\documentclass[twocolumn,prl,floatfix]{revtex4}
\usepackage{graphicx}
\usepackage{amssymb,amsmath,latexsym}
\usepackage{bm}
\usepackage[dvips]{epsfig}

\renewcommand{\vec}[1]{\bm{\mathrm{#1}}}
\newcommand{\vhat}[1]{\hat{\bm{\mathrm{#1}}}}

\begin{document}

\title{Prediction of Giant Spin Motive Force due to Rashba Spin-Orbit Coupling}
\author{Kyoung-Whan Kim$^1$, Jung-Hwan Moon$^2$, Kyung-Jin Lee$^2$, and Hyun-Woo Lee$^1$}
\affiliation{$^1$PCTP and Department of Physics, Pohang University of Science
and Technology, Pohang, 790-784, Korea\\$^2$Department of Materials Science
and Engineering, Korea University, Seoul, 136-701, Korea
}

\date{\today}
%

\begin{abstract}
Magnetization dynamics in a ferromagnet can induce a
spin-dependent electric field through spin motive force. Spin
current generated by the spin-dependent electric field can in turn
modify the magnetization dynamics through spin-transfer torque.
While this feedback effect is usually weak and thus ignored, we
predict that in Rashba spin-orbit coupling systems with large
Rashba parameter $\alpha_{\rm R}$, the coupling generates the
spin-dependent electric field [$\pm(\alpha_{\rm R}m_e/e\hbar)
(\vhat{z}\times
\partial \vec{m}/\partial t)]$, which can be large enough to
modify the magnetization dynamics significantly. This effect
should be relevant for device applications based on ultrathin
magnetic layers with strong Rashba spin-orbit coupling.
\end{abstract}


\maketitle

Similar to the mutual induction between electric and magnetic
fields through the Faraday and Maxwell's laws, spin current and
magnetization induce the dynamics of each other in magnetic
systems through spin-transfer torque
(STT)~\cite{Slonczewski96JMMM,Berger96PRB,Kiselev03Nature,Lee04NM}
and spin motive force
(SMF)~\cite{Berger86PRB,Volovik87JPC,Barnes07PRL}. Through many
years of extensive study~\cite{Kiselev03Nature,Lee04NM}, it has
been demonstrated that STT is a powerful tool to induce the
magnetization dynamics in ferromagnetic nanostructures. In
contrast, SMF has received much less
attention~\cite{Ohe09APL,Yang09PRL,Yamane11PRL} since SMF is very
weak and thus inefficient to generate the spin current.

In this Letter, we demonstrate that SMF can be orders of magnitude
enhanced in systems with strong Rashba spin-orbit coupling (RSOC).
RSOC arises generically when structural inversion symmetry is
broken~\cite{Bychkov84JETPL}. Recent experiments on ultrathin
($\sim$ 1 nm) ferromagnetic layers with the strong structural
inversion asymmetry (such as
Pt/Co/AlO$_x$~\cite{Miron10NatureMat,Pi10APL})
observed large
effective magnetic fields predicted by RSOC
theories~\cite{Obata08PRB,Manchon08PRB}.
A thin ferromagnetic layer in contact with topological
insulators~\cite{Bahramy11preprint,Ishizaka11NM} may also have strong RSOC.
In such magnetic systems, magnetization dynamics can induce large spin
current through SMF, and this spin current can in turn modify the
magnetization dynamics through STT. Thus even for {\it purely
magnetic-field-driven} magnetization dynamics, STT can have sizable magnitude
because of large spin current generated through SMF.
We also propose a
SMF-based method to measure the strength of RSOC in magnetic
systems. This method allows for unambiguous distinction between
RSOC effects and other effects~\cite{Liu11preprint}, which may be
difficult to distinguish through other methods. Thus SMF can also be
a useful tool to quantify RSOC.

RSOC can be described by the Rashba Hamiltonian $H_{\rm
R}$~\cite{Bychkov84JETPL},
\begin{equation}
H_{\rm R} = \frac{\alpha_{\rm
R}}{\hbar}(\vec{\sigma}\times\vec{p})\cdot\vhat{z}
=\frac{\alpha_{\rm R}}{\hbar}(\sigma_xp_y-\sigma_yp_x), \label{HR}
\end{equation}
where the vectors $\vec{\sigma}$ and $\vec{p}$ are the Pauli
matrix and the momentum, respectively. $\vhat{z}$ is the unit
vector along the inversion symmetry breaking direction
(perpendicular to ferromagnetic layer), and $\alpha_{\rm R}$ is
the Rashba constant.
The total Hamiltonian $H$ of a conduction electron then becomes
$H_0+H_{\rm R}$, where $H_0=\vec{p}^2/2m_e-J_{\rm ex}
\vec{\sigma}\cdot\vec{m}$.
Here $m_e$ is effective mass of conduction electrons, $J_{\rm ex}$
($<0$) is the exchange coupling energy,
and $\vec{m}$ is the unit
vector along the local magnetization, which is in general time-
and position-dependent.
An insight into the RSOC effect on SMF can be gained from the
velocity operator
$\vec{v}=[\vec{r},H]/i\hbar=\vec{p}/m_e+\vec{v}_{\rm an}$, where
the anomalous velocity $\vec{v}_{\rm an}=[\vec{r},H_{\rm
R}]/i\hbar=\alpha_{\rm R} \vhat{z}\times \vec{\sigma}/\hbar$
arises due to RSOC.
When the exchange energy is sufficiently larger than RSOC,
$\vec{\sigma}$ for majority (minority) electrons will be
almost anti-parallel (parallel) to $\vec{m}$.
It is then evident that the magnetization
dynamics $\partial\vec{m}/\partial t$ induces the acceleration
$d\vec{v}_{\rm an}/dt$ and the effective electric field $-(m_e/e)
d\vec{v}_{\rm an}/dt$. This heuristic calculation results in the
RSOC contribution to the spin-dependent electric field
$\vec{E}_\pm^{\rm RSOC}$,
\begin{equation}
\vec{E}_\pm^{\rm RSOC}=\pm \alpha_{\rm R} \frac{m_e}{e\hbar}\left(
\vhat{z}\times \frac{\partial \vec{m}}{\partial t} \right),
\label{dEs}
\end{equation}
where $\pm$ applies to the majority and minority electrons,
respectively. The total spin-dependent electric field
$\vec{E}'_\pm$ then becomes $\vec{E}_\pm+\vec{E}_\pm^{\rm RSOC}$,
where
\begin{equation}
\vec{E}_\pm=\pm \sum_i \vhat{x}_i \frac{\hbar}{2e}\left( \frac{\partial
\vec{m}}{\partial t} \times \frac{\partial \vec{m}}{\partial x_i}
\right) \cdot \vec{m} \label{Es0}
\end{equation}
represents the conventional spin-dependent electric field examined
in previous literatures~\cite{Barnes07PRL,Zhang09PRL} in the
absence of RSOC.
%
%
For more rigorous derivation of Eq.~(\ref{dEs}), we may adopt the
calculation scheme in Ref.~\cite{Volovik87JPC}. One first
diagonalizes the exchange coupling term $-J_{\rm
ex}\vec{\sigma}\cdot \vec{m}$ by introducing the unitary operator
$U^\dagger=e^{i\theta\sigma_y/2}e^{i\phi\sigma_z/2}$, where the
Euler angles $(\theta,\phi)$ are defined by spatio-temporal
profile of
$\vec{m}=(\sin\theta\cos\phi,\sin\theta\sin\phi,\cos\theta)$. The
transformed Hamiltonian $H'\equiv U^\dagger HU-i\hbar
U^\dagger\partial_tU$ becomes
$H' =(\vec{p}+e\vec{A}')^2/2m_e-J_{\rm ex}\sigma_z-eA_0'$,
where $\vec{m}$ is rotated to $(0,0,1)$ and the exchange coupling
term is diagonal, $-J_{\rm ex}\sigma_z$. Information on SMF is now
stored in vector and scalar potentials $\vec{A}'$ and $A_0'$ given
by $\vec{A}'=-(i\hbar/e)U^\dagger\nabla U -(\alpha_{\rm
R}m_e/e\hbar)U^\dagger(\vec{\sigma}\times\vhat{z})U$ and
$A_0'=(i\hbar/e)U^\dagger\partial_tU +\alpha_{\rm
R}^2m_e/e\hbar^2$.
Recalling that spins tend to be parallel or anti-parallel to
$\vec{m}=(0,0,1)$,
we may retain only diagonal components of the potentials for
leading order calculation. Then the conventional formula
$-\partial_t \vec{A}'-\nabla A_0'$ determines the total electric
field $\vec{E}'_+$ ($\vec{E}'_-$) for the majority $\sigma_z=-1$
(minority $\sigma_z=+1$) electrons. This calculation reproduces
Eqs.~(\ref{dEs}) and (\ref{Es0}).

We estimate the relative magnitude of $\vec{E}_\pm^{\rm RSOC}$
with respect to $\vec{E}_\pm$; $\vec{E}_\pm^{\rm RSOC}$ is of the
order of $(\alpha_{\rm R} m_e/e\hbar)\omega$, where $\omega$ is
the characteristic angular frequency of the magnetization
dynamics, and $\vec{E}_\pm$ is of the order of
$(\hbar/e)(\omega/L)$, where $L$ is the characteristic length of a
magnetic structure [such as domain wall (DW) width] producing
$\vec{E}_\pm$.
The relative ratio becomes $\alpha_{\rm R} m_e L / \hbar^2 \approx
\alpha_{\rm R} L \times 10^{38}$ (kg/J$^2 \cdot$ s$^2$) for
$m_e=9.1\times 10^{-31}$ kg.
For the value $\alpha_{\rm R}= 10^{-10}$ eV$\cdot$m reported for
Pt/Co(0.6 nm)/AlO$_x$~\cite{Miron10NatureMat} and for $L=$ 20 nm,
this ratio becomes about 30. Thus $\vec{E}_\pm^{\rm RSOC}$ can be
more than an order of magnitude larger than $\vec{E}_\pm$.

\begin{figure}
\includegraphics[width=8.6cm]{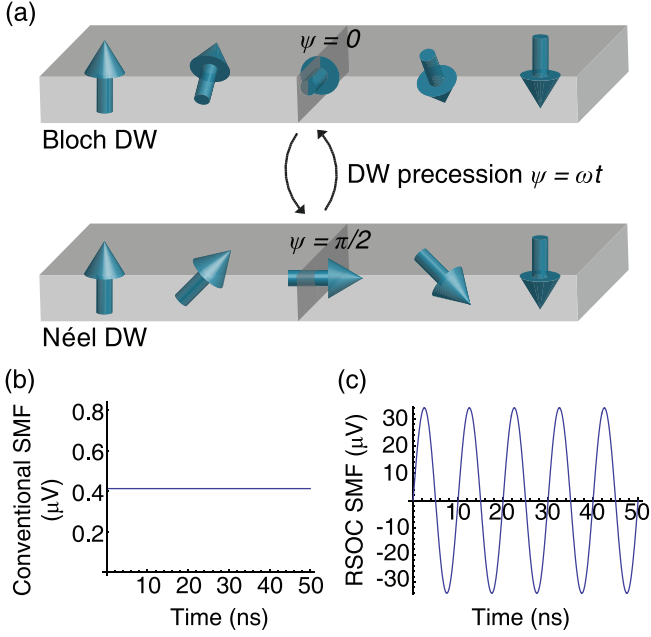}
\caption{(color online) Voltage produced by a precessing DW in a
nanowire. For simplicity, the DW motion along the nanowire is
assumed to be suppressed by a notch in the nanowire. (a) Schematic structure of DW.
DW is assumed to precess and change its structure periodically
between the Bloch DW
 ($\psi=0$ or $\pi$) and the Neel DW
($\psi=\pi/2$ or $3\pi/2$) with $\omega=d\psi/dt=2\pi\times$100
MHz.   (b) Voltage between the two ends of the nanowire due to
$\vec{E}_\pm$. This voltage does not depend on time $t$ in this
particular case since the $t$-dependences of $\vec{m}$, $\partial
\vec{m}/\partial t$, and $\partial \vec{m}/\partial x_i$ mutually
cancel in Eq.~(\ref{Es0}). The voltage magnitude is comparable to
the value reported in Ref.~\cite{Yang09PRL}. (c) Voltage due to
$\vec{E}_\pm^{\rm RSOC}$, which oscillates with $t$. Here
$\alpha_{\rm R}=10^{-10}$ eV$\cdot$m~\cite{Miron10NatureMat},
$P=1$, and the DW width of 20 nm are assumed.
}
\label{Fig:DWM}
\end{figure}


For magnetic materials with nonzero spin polarization $P$,
spin-dependent electric fields can be measured through conventional electric voltage measurement~\cite{Ohe09APL}.
Figure~\ref{Fig:DWM} compares electrical voltages produced by a precessing DW in a nanowire.
%
The voltage due to $\vec{E}_\pm^{\rm RSOC}$ [Fig.~\ref{Fig:DWM}(c)]
exhibits time-dependence different from that due to $\vec{E}_\pm$ [Fig.~\ref{Fig:DWM}(b)]
because of the functional form difference between Eqs.~(\ref{dEs}) and (\ref{Es0}).
Note that the oscillation amplitude in Fig.~\ref{Fig:DWM}(c) is about 80 times larger than
the dc value in Fig.~\ref{Fig:DWM}(b).

In regard to the value of $\alpha_{\rm R}$, $10^{-10}$ eV$\cdot$m
for Pt/Co(0.6 nm)/AlO$_x$~\cite{Miron10NatureMat} is not
exceptional. The $\alpha_{\rm R}$ value in the range
$(0.4-3)\times 10^{-10}$ eV$\cdot$m was reported by photoelectron
spectroscopy measurements~\cite{Henk04JP,Ast07PRL} for a thin
nonmagnetic metal layer in contact with a heavy atomic element
layer. Since magnetism does not suppress $\alpha_{\rm R}$, its
value can be in a similar range for proper combinations of a thin
magnetic layer in contact with a heavy atomic element layer.
A recent effective field measurement result~\cite{Suzuki11APL} for
Ta/CoFeB(1 nm)/MgO implies $\alpha_{\rm R}\approx 0.2 \times
10^{-10}$ eV$\cdot$m,
which 
falls close to the common range.
For this particular multilayer, it was
suggested~\cite{Suzuki11APL} that the contact with the oxide layer
MgO may also be an important source of $\alpha_{\rm R}$. Still
another candidate system is a thin magnetic layer in contact with
topological insulators. For instance, BiTeI has a large bulk RSOC
with $\alpha_{\rm R}=3.8\times 10^{-10}$
eV$\cdot$m~\cite{Ishizaka11NM} and was
predicted~\cite{Bahramy11preprint} to become a topological
insulator under pressure.
%

Large spin-dependent electric field implies large spin current.
The spin current density generated by SMF becomes
$\vec{J}_{\rm s}^{\rm SMF} =
\sigma_\uparrow (\vec{E}_+ + \vec{E}_+^{\rm RSOC})-\sigma_\downarrow
(\vec{E}_- + \vec{E}_-^{\rm RSOC})=\sigma_{\rm c}(\vec{E}_+ + \vec{E}_+^{\rm RSOC})$~\cite{spin-current-density},
where $\sigma_{\uparrow(\downarrow)}$ is the longitudinal electrical conductivity of
majority (minority) electrons and $\sigma_{\rm
c}=\sigma_\uparrow+\sigma_\downarrow$.
Since $\vec{E}_+^{\rm RSOC}$ is much larger than $\vec{E}_+$
and $\vec{E}_\pm^{\rm RSOC}$ is perpendicular to $\hat{\vec{z}}$ [Eq.~(\ref{dEs})],
$\vec{J}_{\rm s}^{\rm SMF}$
flows {\it within} a magnetic layer.
Then for a typical in-plane charge conductivity $\sigma_{\rm
c}=10^7 \ \Omega^{-1}\cdot \textrm{m}^{-1}$ of metallic
ferromagnetic layers and $\alpha_{\rm R}=(0.2-3)\times 10^{-10}$
eV$\cdot$m, $\vec{J}_{\rm s}^{\rm SMF}$ has magnitude
$(1.8-27)\tau^{-1}$ A$\cdot$s/m$^2$, where $\tau$ is the
characteristic time scale of magnetization dynamics. For fast DW
motion with speed 400 m/s~\cite{Miron11NM}, we find that
$|\vec{J}_{\rm s}^{\rm SMF}|$ goes up to $(0.18-2.7)\times
10^{11}$ A/m$^2$ near the DW center of width 20 nm. Thus RSOC
allows SMF to generate large spin current density for fast
magnetization dynamics.

Large $\vec{J}_{\rm s}^{\rm SMF}$ generated by magnetization
dynamics implies that magnetization dynamics \textit{itself} can
be significantly modified by $\vec{J}_{\rm s}^{\rm SMF}$ through
STT. Next we examine this feedback effect. In conventional
situations where SMF is negligible, the magnetization dynamics is
described by the Landau-Lifshitz-Gilbert (LLG) equation, $\partial
\vec{m}/\partial t = -\gamma\vec{m}\times \vec{H}_{\rm
eff}+\alpha_{\rm G} \vec{m}\times \partial \vec{m}/\partial t +
\vec{T}(\vec{J}_{\rm s})$, where $\vec{T}$ represents the STT and
depends on externally supplied spin current density $\vec{J}_{\rm
s}$. Here $\gamma$ is the gyromagnetic ratio, $\alpha_{\rm G}$ is
the Gilbert damping parameter, and $\vec{H}_{\rm eff}$ is the sum
of an external magnetic field and effective magnetic fields due to
magnetic anisotropy and magnetic exchange energy.
To examine the feedback effect, 
one simply needs to replace $\vec{T}(\vec{J}_{\rm s})$ by $\vec{T}(\vec{J}_{\rm s}+\vec{J}_{\rm s}^{\rm SMF})
=\vec{T}(\vec{J}_{\rm s})+\vec{T}(\vec{J}_{\rm s}^{\rm SMF})$
to obtain the modified LLG equation,
\begin{equation}
\frac{\partial \vec{m}}{\partial t} = -\gamma\vec{m}\times
\vec{H}_{\rm eff}+\alpha_{\rm G} \vec{m}\times \frac{\partial
\vec{m}}{\partial t}
 + \vec{T}(\vec{J}_{\rm s})+\vec{T}(\vec{J}_{\rm s}^{\rm SMF}).
 \label{eq:modified-LLG}
\end{equation}
Note that STT $\vec{T}$ now has two spin current sources, $\vec{J}_{\rm s}$ and $\vec{J}_{\rm s}^{\rm SMF}$.
Thus even when there is no externally supplied spin current ($\vec{J}_{\rm s}=0$),
STT still affects the magnetization dynamics
as long as $\vec{J}_{\rm s}^{\rm SMF}$ is not zero.
Therefore STT becomes relevant even for {\it purely field-driven} magnetization dynamics.

To gain an insight into roles of the feedback STT $\vec{T}(\vec{J}_{\rm s}^{\rm SMF})$,
we express it in the following form,
\begin{equation}
\vec{T}(\vec{J}_{\rm s}^{\rm SMF})=\vec{m}\times \mathcal{D}\cdot \frac{\partial\vec{m}}{\partial t}.
\label{eq:feedback-STT}
\end{equation}
This form is natural since $\vec{T}$ is orthogonal to $\vec{m}$
[thus $\vec{T}(\vec{J}_{\rm s}^{\rm SMF})=\vec{m}\times$ (function
of $\vec{J}_{\rm s}^{\rm SMF}$)] and $\vec{J}_{\rm s}^{\rm SMF}$
is proportional to $\partial \vec{m}/\partial t$ [thus (function
of $\vec{J}_{\rm s}^{\rm SMF}$) $= \mathcal{D}\cdot \partial
\vec{m}/\partial t$]. Here $\mathcal{D}$ will depend on $\vec{m}$
and be a $3\times 3$ matrix in general. Although $\mathcal{D}$ is
not a constant, the structural similarity between
Eq.~(\ref{eq:feedback-STT}) and the Gilbert damping torque
$\alpha_{\rm G} \vec{m}\times \partial \vec{m}/\partial t$ implies
that $\mathcal{D}$ may be interpreted as a correction to the
Gilbert damping parameter $\alpha_{\rm G}$. Thus roles of the
feedback STT can be analyzed by using this damping correction
picture.

After explicit calculation, we find that the matrix elements of the $3\times 3$ matrix $\mathcal{D}$ are given by
\begin{equation}
\mathcal{D}_{ij} = \eta\sum_k \left( X_{ki} + \tilde{\alpha}_{\rm
R} \epsilon_{3ki} \right) \left( X_{kj} + \tilde{\alpha}_{\rm R}
\epsilon_{3kj} \right), \label{eq:Damping-tensor}
\end{equation}
where $X_{ki} = \left( \vec{m}\times \partial \vec{m}/\partial
x_k\right)_i$, $\eta=\mu_{\rm B} \hbar \sigma_{\rm c}/2e^2 M_{\rm
s}$, $\tilde{\alpha}_{\rm R} = 2\alpha_{\rm R} m_e/\hbar^2$,
$\mu_{\rm B} (>0)$ is the Bohr magneton, $M_{\rm s}$ is the
saturation magnetization, and $\epsilon_{ijk}$ is the Levi-Civita
permutation symbol ($\epsilon_{123}=1$).
To derive Eq.~(\ref{eq:Damping-tensor}), we simply combine the
spin-dependent electric field equations~[Eqs.~(\ref{dEs}) and
(\ref{Es0})] with known contributions to STT. To be specific, the
adiabatic STT contribution [$\propto (\vec{J}_{\rm s}\cdot
\nabla)\vec{m}$] and the field-like STT contribution [$\propto
\alpha_{\rm R} \vec{m}\times (\hat{\vec{z}} \times \vec{J}_{\rm
s})$]~\cite{Obata08PRB,Manchon08PRB} are taken into account in the
calculation whereas the nonadiabatic STT contribution and recently
discovered Slonczewski-like STT
contribution~\cite{Wang11preprint,Kim11preprint,Pesin12preprint}
are ignored since the latter two contributions are smaller than
the former two.

The importance of the feedback STT can be estimated from the
relative magnitude of $\mathcal{D}$ with respect to $\alpha_{\rm
G}$. Here $\alpha_{\rm G}$ includes all contributions other than
the SMF contribution. The intrinsic bulk Gilbert damping parameter
is of the order of 0.01 for typical metallic ferromagnets. In a
thin magnetic layer, this intrinsic value is enhanced by the
conventional spin pumping
mechanism~\cite{Tserkovnyak02PRL,comment-SPvsRSOC} and the
enhanced value is estimated to $0.1/d_{\rm F}$ (nm), which is 0.1
for thin magnetic layer of thickness $d_{\rm F}=1$ nm. Thus for
the feedback effect to be a relevant factor for the magnetization
dynamics of a thin magnetic layer, $\mathcal{D}$ should be
comparable to or larger than 0.1.
When RSOC is absent ($\tilde{\alpha}_{\rm R}=0$), $\mathcal{D}$
reduces to the result in Ref.~\cite{Zhang09PRL} and is of the
order of $\eta/L^2$. For $\sigma_{\rm c}=10^{7}\
\Omega^{-1}\cdot\textrm{m}^{-1}$, $M_{\rm s}=10^6$ A/m and $L=20$
nm, $\eta$ is 0.2 nm$^2$ and $\eta/L^2$ is 0.0005. Thus except for
special situations with $L\lesssim$ 1 nm, $\mathcal{D}$ becomes
much smaller than 0.1 and the feedback effect is negligible.
%
When RSOC is strong, on the other hand, the largest contribution
to $\mathcal{D}$ arises from the term $\eta \tilde{\alpha}_{\rm
R}^2\sum_k \epsilon_{3ki}\epsilon_{3kj}$ and thus $\mathcal{D}$
becomes of the order of $\alpha_{\rm G}'\equiv \eta
\tilde{\alpha}_{\rm R}^2$~\cite{Hankiewicz07PRB}. For $\alpha_{\rm
R}=10^{-10}$ eV$\cdot$ m~\cite{Miron10NatureMat},
$\tilde{\alpha}_{\rm R}$ is $2.6$ nm$^{-1}$ and  $\alpha_{\rm G}'$
is $1.4$. Since $\alpha_{\rm G}'\gg \alpha_{\rm G}$, one of
evident effects of the feedback effect is to enhance the effective
damping significantly.
%

For deeper understanding of the feedback effect, however, one
should
address the detailed structure of the damping matrix $\mathcal{D}$
in Eq.~(\ref{eq:Damping-tensor}). Extensive discussion on
implications of the damping matrix structure will be presented
elsewhere. Here we present one simple implication contained in the
largest contribution $\alpha_{\rm G}'\sum_k
\epsilon_{3ki}\epsilon_{3kj}$, which reduces to $\alpha_{\rm
G}'(\delta_{ij}-\delta_{3i}\delta_{3j})$ after simple algebra.
This contribution makes the effective damping anisotropic
since  $\sum_j(\delta_{ij}-\delta_{3i}\delta_{3j})(\partial \vec{m}/\partial t)_j$
becomes $(\partial \vec{m}/\partial t)_i$
when $\partial \vec{m}/\partial t$ is perpendicular to $\hat{z}$-direction,
and vanishes when $\partial \vec{m}/\partial t$ is parallel to $\hat{z}$-direction.
For a magnetic layer with the perpendicular magnetic anisotropy,
this damping anisotropy can be tested in experiments by measuring
the effective damping in two different ways. When the effective
damping is measured by ferromagnetic resonance, $\partial
\vec{m}/\partial t$ remains essentially perpendicular to
$\hat{z}$-direction and thus the effective damping becomes
$\alpha_{\rm G}+\alpha_{\rm G}'$. The effective damping may be
measured alternatively by using the relation that the field-driven
DW velocity below the Walker-breakdown threshold is inversely
proportional to the effective damping. This measurement should
produce a smaller effective damping value than the ferromagnetic
resonance since $\partial \vec{m}/\partial t$ is along the
$\pm\hat{z}$-direction at the DW center.
From a simple calculation based on the Thiele's collective
coordinate description~\cite{Thiele73PRL} of the DW configuration
in terms of the DW position and DW tilting angle, we find that the
effective damping becomes $\alpha_{\rm G}+\alpha_{\rm G}'/3$.
Thus in strong RSOC systems, where $\alpha_{\rm G}'\gg \alpha_{\rm
G}$, we predict that the effective damping for the ferromagnetic
resonance is about factor 3 larger than that for the DW motion. By
the way, Ref.~\cite{Miron10NatureMat} determined the effective
damping for Pt/Co(0.6 nm)/AlO$_x$ from the field-driven DW motion
measurement and found 0.5, which is in reasonable agreement with
our prediction $\alpha_{\rm G}+\alpha_{\rm G}'/3\approx
\alpha_{\rm G}'/3\approx 0.5$ for the system.

\begin{figure}
\includegraphics[width=8.6cm]{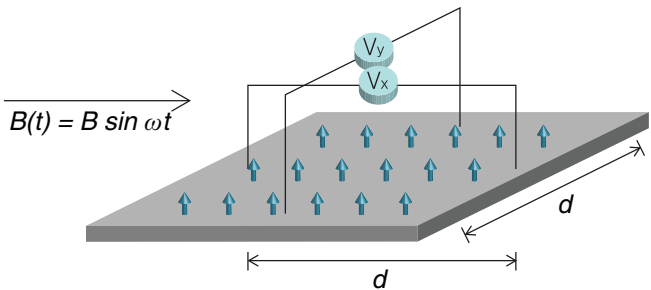}
\caption{(color online)
Measurement scheme of the RSOC effect on the SMF for a uniformly magnetized layer with the perpendicular magnetic anisotropy.
%
} \label{fig:rashba}
\end{figure}

Figure~\ref{fig:rashba} shows schematically an experimental setup
to test the RSOC-enhanced SMF directly. When a {\it uniformly}
magnetized magnetic domain precesses due to external magnetic
field $B(t)=B\sin \omega t$,
$\vec{E}_\pm^{\rm RSOC}$ will generate electric voltages
$V_x$ and $V_y$.
Since $\vec{E}_\pm^{\rm RSOC}$ is position-independent [Eq.~(\ref{dEs})],
these voltages should be proportional to
the spacing $d$ between electrodes.
Also since the direction of $\vec{E}_\pm^{\rm RSOC}$ rotates within $xy$-plane
as $\partial \vec{m}/\partial t$ rotates within $xy$-plane (due to the perpendicular magnetic anisotropy),
the voltages will be oscillatory with the oscillation phases of $V_x$ and $V_y$
different from each other by about 90$^\circ$.
These features are qualitatively different from other effects that may affect this measurement.
The conventional spin pumping~\cite{Tserkovnyak02PRL} may also contribute to the
electric voltages if the two electrical contacts with the electrodes
are not identical, as demonstrated in Ref.~\cite{Costache06PRL}.
These contribution is however independent of $d$ and thus distinguishable from the RSOC-enhanced SMF contribution.
Spin Hall effect in contacts or the neighboring heavy metal layer can also generate a contribution.
This spin Hall contribution is however different from the RSOC-enhanced SMF contribution
since $\hat{\vec{z}}\times \vec{m}$ determines the direction of the voltage gradient in case of the spin Hall contribution
whereas $\hat{\vec{z}}\times \partial\vec{m}/\partial t$ does in case of the RSOC-enhanced SMF contribution.
Recalling that $\vec{m}$ and $\partial \vec{m}/\partial t$ are orthogonal to each other,
these two contributions differ by the oscillation phase of 90$^\circ$.
A recent experiment~\cite{Liu11preprint} demonstrated that in
certain experimental situations~\cite{Miron11Nature}, the spin
Hall effect can be very similar to the RSOC effect.
The electrical voltage measurement in Fig.~\ref{fig:rashba} can be
a very decisive experiment to verify the RSOC-enhanced SMF and
also RSOC itself free from such ambiguities. The Rashba constant
$\alpha_{\rm R}$ can be determined from the proportionality
constant between the voltages and $d$.
%
%

Lastly two remarks are in order. Firstly, temperature may affect
the predicted phenomena. The value of $\alpha_{\rm R}$ may depend
on temperature. But recalling that the reported value of
$\alpha_{\rm R}=10^{-10}$ eV$\cdot$m for
Pt/Co/AlO$_x$~\cite{Miron10NatureMat} was obtained at room
temperature, we expect that $\alpha_{\rm R}$ can stay large even
at room temperature. Thermal fluctuation of $\vec{m}$ is another
possible source of the temperature dependence of SMF. Since the
fluctuation is not correlated in space and time, it will not
affect the SMF voltage measurements in Figs.~\ref{Fig:DWM} and
\ref{fig:rashba} significantly.
Thus we expect that temperature effects do not modify SMF effects
qualitatively.
Secondly, though not elaborated here, RSOC gives rise to not only
$\vec{E}_\pm^{\rm RSOC}$ [Eq.~(\ref{dEs})] but also a
spin-dependent magnetic field. Our preliminary calculation
indicates that certain types of magnetization dynamics such as DW
motion may be accelerated by this effective magnetic field.
Further research is required to understand its effects.

In conclusion, we demonstrated that SMF can be orders of
magnitudes enhanced in strong RSOC systems. The enhanced SMF is
strong enough to modify the magnetization dynamics significantly.
RSOC-enhanced SMF may affect performance of various device
applications based on ultrathin magnetic
layers~\cite{Ikeda10NM,Miron11Nature,Miron11NM} since RSOC effects
tend to get larger in thinner systems. Thus SMF is not a weak
effect of purely scientific concern any more but instead a
technologically relevant effect that should be taken into account
in future studies.
As a final remark, during the revision of this manuscript, we were informed that
two other groups~\cite{Kohno,Maekawa} also obtained Eq.~(\ref{dEs}).

\begin{acknowledgements}
We gratefully acknowledge M. Stiles for critical comments on the manuscript.
We also acknowledge B.-C. Min, S.-Y. Park, and Y. Jo for stimulating
discussion. This work was financially supported by the NRF (2009-0084542,
2010-0014109, 2010-0023798, 2011-0028163, 2011-0030784) and BK21. KWK
acknowledges the financial support by the NRF (2011-0009278) and TJ Park.
\end{acknowledgements}

\end{document}